\documentclass[twocolumn, floatfix]{aastex631} 
\usepackage{natbib}
\usepackage{xspace}
\usepackage{xcolor}
\usepackage{graphicx}
\usepackage{hyperref}
\usepackage{amsmath}
\usepackage{xcolor}
\usepackage{soul}
\usepackage{multirow}
\usepackage[normalem]{ulem}
\usepackage{soul}

\hypersetup{linkcolor=red,citecolor=cyan,filecolor=magenta,urlcolor=blue}

\newcommand{\lowZlowR}{$0.86^{+0.09}_{-0.10}$ }
\newcommand{\lowZhighR}{$0.40^{+0.13}_{-0.08}$ }

\newcommand{\lowZgroups}{0.82$^{+0.09}_{-0.10}$ }
\newcommand{\highZgroups}{0.59$^{+0.13}_{-0.08}$ }
\newcommand{\lowZfield}{$0.59^{+0.11}_{-0.10}$}
\newcommand{\highZfield}{$0.55^{+0.08}_{-0.08}$}
\newcommand{\lowZfieldAve}{$0.52^{+0.08}_{-0.08}$}

\begin{document}

\title{Investigating the Dominant Environmental Quenching Process in UVCANDELS/COSMOS Groups}

\author[0000-0002-7830-363X]{Maxwell Kuschel}
\affiliation{Minnesota Institute of Astrophysics and School of Physics and Astronomy, University of Minnesota, Minneapolis, MN, USA}


\author[0000-0002-9136-8876]{Claudia Scarlata} 
\affiliation{Minnesota Institute of Astrophysics and School of Physics and Astronomy, University of Minnesota, Minneapolis, MN, USA}

\author[0000-0001-7166-6035]{Vihang Mehta} 
\affiliation{IPAC/Caltech, Pasadena, CA, USA}

\author[0000-0002-7064-5424]{Harry I. Teplitz}
\affiliation{IPAC, Mail Code 314-6, California Institute of Technology, 1200 E. California Blvd., Pasadena CA, 91125, USA} 

\author[0000-0002-9946-4731]{Marc Rafelski}
\affiliation{Space Telescope Science Institute, Baltimore, MD, USA}
\affiliation{Department of Physics and Astronomy, Johns Hopkins University, Baltimore, MD 21218, USA} 

\author[0000-0002-9373-3865]{Xin Wang}
\affiliation{IPAC/Caltech, Pasadena, CA, USA} 

\author[0000-0003-3759-8707]{Ben Sunnquist}
\affiliation{Space Telescope Science Institute, Baltimore, MD, USA} 

\author[0000-0002-0604-654X]{Laura Prichard}
\affiliation{Space Telescope Science Institute, Baltimore, MD, USA} 

\author[0000-0001-9440-8872]{Norman Grogin}
\affiliation{Space Telescope Science Institute, Baltimore, MD, USA} 

\author[0000-0002-6610-2048]{Anton Koekemoer}
\affiliation{Space Telescope Science Institute, Baltimore, MD, USA} 

\author[0000-0001-8156-6281]{Rogier Windhorst}
\affiliation{Arizona State University, Tempe, AZ, USA} 

\author[0000-0001-7016-5220]{Michael Rutkowski}
\affiliation{Minnesota State University-Mankato, Mankato, MN, USA} 


\author[0000-0002-8630-6435]{Anahita Alavi}
\affiliation{IPAC/Caltech, Pasadena, CA, USA}

\author[0000-0003-3691-937X]{Nima Chartab} 
\affiliation{The Observatories of the Carnegie Institution for Science, 813 Santa Barbara St., Pasadena, CA 91101, USA}

\author[0000-0003-1949-7638]{Christopher J. Conselice}
\affiliation{University of Manchester, Manchester, UK} 

\author[0000-0002-7928-416X]{Y.Sophia Dai}
\affiliation{Chinese Academy of Sciences South America Center for Astronomy (CASSACA), National Astronomical Observatories of China (NAOC), 20A Datun Road, Beijing, 100012, China} 

\author[0000-0003-1530-8713]{Eric Gawiser}
\affiliation{Rutgers University, New Brunswick, NJ, USA} 

\author[0000-0002-7831-8751]{Mauro Giavalisco}
\affiliation{University of Massachusetts, Amherst, MA, USA} 

\author[0000-0002-7959-8783]{Pablo Arrabal Haro}
\affiliation{NSF's National Optical-Infrared Astronomy Research Laboratory, Tucson, AZ, USA} 

\author[0000-0001-6145-5090]{Nimish Hathi}
\affiliation{Space Telescope Science Institute, Baltimore, MD, USA} 

\author[0000-0003-1268-5230]{Rolf A.~Jansen}
\affiliation{Arizona State University, Tempe, AZ, USA} 

\author[0000-0001-7673-2257]{Zhiyuan Ji}
\affiliation{University of Massachusetts, Amherst, MA, USA} 

\author[0000-0003-1581-7825]{Ray A. Lucas}
\affiliation{Space Telescope Science Institute, Baltimore, MD, USA} 

\author{Kameswara Mantha} 
\affiliation{Minnesota Institute of Astrophysics and School of Physics and Astronomy, University of Minnesota, Minneapolis, MN, USA}

\author{Bahram Mobasher}
\affiliation{University of California, Riverside, CA, USA} 

\author{Robert W. O'Connell}
\affiliation{University of Virginia, Charlottesville, VA, USA} 

\author[0000-0002-4271-0364]{Brant Robertson}
\affiliation{University of California, Santa Cruz, Santa Cruz, CA 95064 USA} 

\author[0000-0002-0364-1159]{Zahra Sattari} 
\affiliation{Department of Physics and Astronomy, University of California, Riverside, 900 University Ave, Riverside, CA 92521, USA}

\author[0000-0003-3466-035X]{L. Y. Aaron Yung}
\affiliation{NASA Goddard Space Flight Center, Greenbelt, MD, USA} 


\author[0000-0003-2842-9434]{Romeel Dav\'e}
\affiliation{Institute for Astronomy, University of Edinburgh, Royal Observatory, Blackford Hill, Edinburgh EH9 3HJ, UK}

\author{Duilia DeMello}
\affiliation{The Catholic University of America, Washington, DC, USA}

\author[0000-0001-5414-5131]{Mark Dickinson}
\affiliation{NSF's National Optical-Infrared Astronomy Research Laboratory, Tucson, AZ, USA}

\author[0000-0001-7113-2738]{Henry Ferguson}
\affiliation{Space Telescope Science Institute, Baltimore, MD, USA}

\author[0000-0001-8519-1130]{Steven L. Finkelstein}
\affiliation{University of Texas, Austin, TX, USA}

\author{Matt Hayes}
\affiliation{Stockholm University, Stockholm, Sweden}

\author[0000-0002-5924-0629]{Justin Howell}
\affiliation{IPAC/Caltech, Pasadena, CA, USA}

\author[0000-0002-5601-575X]{Sugata Kaviraj}
\affiliation{University of Hertfordshire, Hertfordshire, UK}

\author[0000-0001-6529-8416]{John W. Mackenty}
\affiliation{Space Telescope Science Institute, Baltimore, MD, USA}

\author[0000-0002-4935-9511]{Brian Siana}
\affiliation{University of California, Riverside, CA, USA}

\date{\today}

\begin{abstract}
     {We explore how the fraction of quenched galaxies changes in groups of galaxies with respect to the distance to the center of the group, redshift, and stellar mass to determine the dominant process of environmental quenching in $0.2 < z < 0.8$ groups. We use new UV data from the  UVCANDELS project in addition to existing multiband photometry to derive new galaxy physical properties of the group galaxies from the zCOSMOS 20k Group Catalog. Limiting our analysis to a complete sample of log$(M_*/M_{\odot})>10.56$ group galaxies we find that the probability of being quenched increases slowly with decreasing redshift, diverging from the stagnant field galaxy population. A corresponding analysis on how the probability of being quenched increases with time within groups suggests that the dominant environmental quenching process is characterized by slow ($\sim$Gyr) timescales. We find a quenching time of approximately $4.91^{+0.91}_{-1.47} $Gyrs, consistent with the slow processes of strangulation \citep{larson_evolution_1980} and delayed-then-rapid quenching \citep{wetzel_galaxy_2013}, although more data are needed to confirm this result.
    }
\end{abstract}

\section{Introduction}
\label{sect:intro}

Star formation in a galaxy requires the cooling of H\textsc{i} gas that eventually collapses and forms stars \citep{jeans_i_1902, oosterloo_h_2001}. As a galaxy  consumes its cold gas content via star formation it eventually leads to cessation of star formation if the cold gas reservoir is not replenished. Galaxies are both observed and predicted to be hosted in halos containing large amounts of warm/hot gas, where cold streams can funnel cold gas into galaxies, slowly fueling star-formation over time \citep{lilly_gas_2013, ford_baryon_2016,werk_cos-halos_2016}. Both in the local universe, and out to redshifts z$\sim$2 galaxies are observed to obey a bimodal distribution in their star-formation rates \citep{brammer_number_2011, muzzin_evolution_2013, peng_mass_2010}. Many galaxies are found to be star-forming at rates according to the star formation rate - stellar mass (SFR-M) Main Sequence, meaning the extent of star-formation is dependent on the galaxy stellar mass and redshift \citep{speagle_highly_2014}. At all times, however, there exists a population of galaxies in which no new stars are being formed, having ceased all significant star-formation at earlier times. These galaxies are typically referred to as quenched, quiescent, or passive galaxies.  Various mechanisms have been proposed to explain why a galaxy stops forming stars. These quenching mechanisms could be internal to the galaxies themselves \citep{martig_morphological_2009,he_properties_2019, murray_radiation_2011} or be the result of interactions between galaxies and the environment in which they reside \citep{peng_mass_2010, knobel_colors_2013, presotto_journey_2012, wetzel_galaxy_2013, jian_pan-starrs1_2017, wagner_evolution_2016}.  

Large (log(M$_*$/M$_{\odot}) > 12$) dark matter halos containing multiple subhalos of galaxies, known as the group and cluster environments, have been shown to influence SFR in both observations and hydrodynamical simulations \citep{donnari_quenched_2020, jian_pan-starrs1_2017, wagner_evolution_2016}. It is theorized that this occurs due to the large relative velocities that galaxies have compared to the intracluster medium in groups and clusters \citep{larson_evolution_1980, gunn_infall_1972}. Furthermore, the group and cluster environments are more densely populated than the field, causing more interactions between galaxies in the form of dynamical friction and galactic mergers.

In this paper we investigate the role the group environment plays in determining a galaxy's probability of being quenched. In what follows, we will refer to ``environmental quenching'' as the ensemble of physical processes that affect the star-formation rate in galaxies, and are a consequence of the fact that a galaxy resides in a subhalo of a more massive halo. We will also work to separate these effects from the quiescent population due to internal processes driving quenching, known as ``mass quenching'', which significantly influences high mass galaxies \citep{wetzel_galaxy_2013, donnari_quenched_2020, jian_pan-starrs1_2017}.

Star-formation is regulated, in part, by the interactions (gravitational and/or hydrodynamic) between the galaxies' gas and dark matter halo and/or gas content inside of the host halo. Tidal- and ram-pressure stripping are very efficient quenching mechanisms, particularly in the cluster environment \citep{mcpartland_jellyfish_2016}. Ram-pressure stripping results from the relative motion of satellites (galaxies hosted together in a larger dark matter halo) inside a gas rich halo \citep{gunn_infall_1972}, while tidal stripping is a consequence of varying tidal forces acting on a satellite as it moves in the gravitational potential of the halo \citep{moore_galaxy_1996}. These mechanisms are particularly efficient in very massive (log(M$_*$/M$_{\odot})>14$), cluster-size, halos and on low-density galaxies, but are also observed in smaller group-size halos \citep{jian_pan-starrs1_2017, larson_infall_1972}. 

Broadly speaking, two main paths are possible for the environmental quenching of star formation in satellites, which are characterized by the associated time-scales. Ram-pressure stripping can influence star-formation by directly removing the galaxy's cold gas (i.e., the interstellar medium) and stripping the outer, lower-density material \citep{gunn_infall_1972}. Ram-pressure stripping is an example of a rapid quenching process, working on timescales of a few million years \citep{taranu_quenching_2014}. Strangulation occurs when only the outer material is stripped, leaving the ISM gas untouched to continue forming stars for several billion years \citep{larson_evolution_1980}. Because much of the ISM is still in place this process is much slower than Ram-Pressure Stripping, quenching on timescales of $\sim 3$~Gyr \citep{kawata_strangulation_2008}. 

We attempt to disentangle the relative importance of fast and slow quenching mechanisms by investigating the probability that a galaxy is quenched as a function of the time it has been in a group. 

We focus on galaxy groups identified in the CANDELS/COSMOS field (Section~\ref{sect:catalog}), using the stellar population properties derived from the new UVCANDELS dataset (Wang et al in prep., Section~\ref{sect:catalog}). We detail our methods of analysis in Section~\ref{sect:analysis}, and present our results in Section~\ref{sect:results}. Comparison with other studies is presented in Section~\ref{sec:discussion}, followed by our conclusion in Section~\ref{sect:con}.
Throughout the paper, we assume a standard $\Lambda CDM$ cosmology with parameters $H_0 = 68.8~km~s^{-1}~Mpc^{-1}$, $\Omega_M = 0.315$ and $\Omega_{\Lambda} = 0.685$ \citep{gray_pixelated_2022, planck_collaboration_planck_2020}, all magnitudes are expressed in the  AB system \citep{oke_absolute_1974}, and where applicable we use the Chabrier IMF \citep{chabrier_galactic_2003}.

\begin{figure}[ht!]
  \begin{center}
    \includegraphics[width=\linewidth]{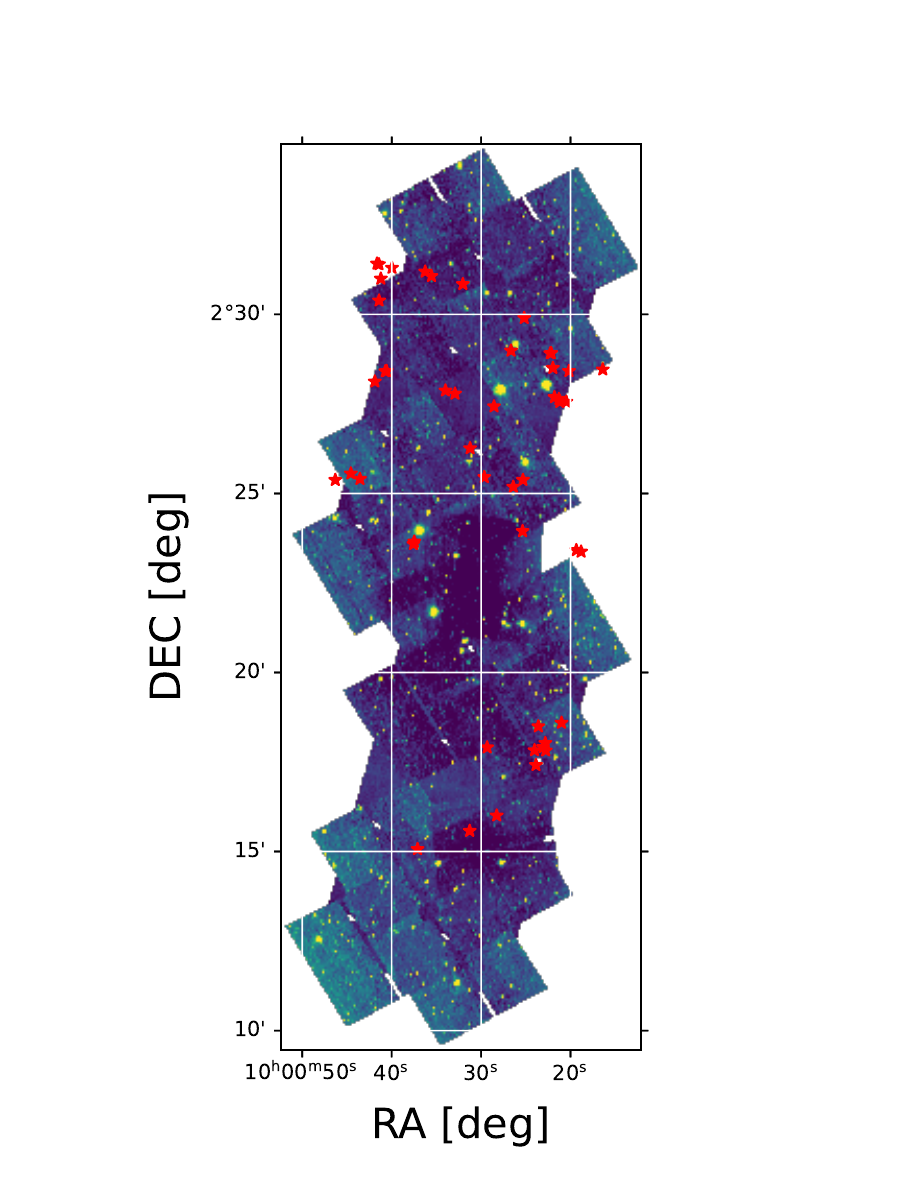}
    \caption{A plot of F275W and F435W mosaics overlayed with the group members being used. All group members have been reprocessed in UVCANDELS as is described in Section~\ref{sec:UVcatalog}, but not all are covered by the F275W or F435W mosaics.}
    \label{fig:Stamps}
  \end{center}
\end{figure}

\section{The Data}
\label{sect:catalog}
In this work we study the fraction of quenched galaxies in groups and in the field. We derive new estimates of the galaxy physical properties using data from the recently completed UVCANDELS Hubble Treasury program (Wang et al. in prep., \citet{koekemoer_candels_2011, grogin_candels_2011}, Section~\ref{sec:UVcatalog}). For the groups analysis we start from the galaxy groups identified in the $z$COSMOS 20k Group Catalog of \citet[][hereafter K12, Section~\ref{sec:groupcatalog}]{knobel_zcosmos_2012}.

\subsection{UVCANDELS Catalog}
\label{sec:UVcatalog}

The area of COSMOS covered with HST optical and NIR observations by the CANDELS program \citep{grogin_candels_2011, koekemoer_candels_2011} was one of the targets for follow-up F275W WFC3/UVIS and F435W ACS/WFC observations with the recent UVCANDELS program which covers 4 of the 5 CANDELS fields (COSMOS, GOODS-N, GOODS-S, AEGIS) (PI: H. Teplitz, Cycle 26 GO 15647, F275W mosaics are currently available at the Mikulski Archive for Space Tele-
scopes at \hyperlink{https://archive.stsci.edu/hlsp/uvcandels}{doi:10.17909/8s31-f778}). The addition of the observed UV and blue optical B-band is particularly important for increasing the accuracy of the photometric redshifts used in this analysis \citep{rafelski_uvudf_2015}.

The physical parameters we use in our analysis are derived by fitting the spectral energy distributions (SEDs) of our sample, obtained using the new F275W and F435W filters in addition to existing photometry (presented in further detail in Mehta et al. in prep). The existing photometry that we use includes the \textit{HST/ACS} F606W and F814W bands as well as the \textit{HST/WFC3} F125W and F160W bands available from \citet{nayyeri_catalog_2017}. We also include the photometry from CFHT/MegaPrime $u^\star$, $g^\star$, $r^\star$, $i^\star$ and $z^\star$, the Subaru/SuprimeCam $B$, $g^+$, $V$, $r^+$, $i^+$ and $z^+$, VLT/VISTA Y, J, H and K, Mayall/NEWFIRM $J1$, $J2$, $J3$, $H1$, $H2$, and $K$ as well as \textit{Spitzer/IRAC} ch. 1, 2, 3 and 4 bands that are also available as part of the \citet{nayyeri_catalog_2017} catalog. The combination of these filters allows for a robust SED fitting, even for galaxies where some filters may be missing, such as some galaxies on the edge of the CANDELS/COSMOS field.

The stellar physical properties used for the analysis presented in this work are computed using CIGALE \citep[Code Investigating GALaxy Emission; ][]{cigale_boquien_19,cigale_burgarella_2005,cigale_noll_2009}. The full description of the SED fitting procedure for the UVCANDELS catalog will be presented in a future publication (Mehta et al. in prep.). Breifly, we use \citet{bruzual_stellar_2003} stellar population models when fitting with a \citet{chabrier_imf_2003} IMF. We allow stellar metallicity to be a free parameter varying between $Z=0.005$ Zsol and 2.5 Zsol. We choose the modified \citet{charlot_fall_2009} dust law to parameterize the dust and allow it to vary as a free parameter between A$_{V,ISM}=0$ and 4 with a $A_{V,ISM}/A_{V,stellar}=0.44$. The stellar formation histories are parameterized as a delayed exponential with e\-folding time is varied over a grid between 30Myr and 30Gyr. Additionally, we allow for the possibility of an episode of recent star-burst as a 10 Myr old burst with an exponential e-folding time of 50 Myr and the contribution of the SF burst is parameterized by the fraction of total mass generated in the burst. When spectroscopic redshifts are available, we did not find a significant difference between the physical properties derived with and without the UV and B-band data. When fitting for group galaxies from K12 we use the group redshifts provided by K12 to match the redshifts used in group identification. These were taken to be the median redshift of the group members, derived from the zCOSMOS survey \citep{lilly_zcosmos_2009}.

For the field galaxies in our sample we use photometric redshifts that have been computed with the inclusion of the new UVCANDELS F275W and F435W photometry which combine redshifts in probability space from multiple redshift codes yielding robusts redshifts. (Sunnquist et al. in prep.). 
When fitting, we add in quadrature a nominal error of 0.02 mag to all photometry in order to account for calibration variance across the various filters.

\subsection{Selection of group galaxy sample}
\label{sec:groupcatalog}
In this work, we rely on the $z$COSMOS 20k Group Catalog presented in K12. K12 use accurate spectroscopic redshifts from the $z$COSMOS  catalog of  \citet{lilly_zcosmos_2009} to identify group galaxies using two common group-finding algorithms, Friends-of-Friends \citep{berlind_percolation_2006, eke_galaxy_2004, huchra_groups_1982} and the Voronoi-Delaunay method \citep{knobel_optical_2009}. These algorithms were calibrated using simulated mock galaxy catalogs extracted from the Millenium~I Dark Matter N-Body simulation \citep{springel_simulations_2005} by \citet{kitzbichler_high-redshift_2007}. The mock catalogs were used to fine-tune the parameters of the group identification algorithms in order to reach a completeness and purity of $\gtrsim 80$\% for groups with 3 or more members. From this K12 catalog we also use the derived group halo-masses and group halo radii. These radii and masses were derived utilizing their mock catalogs by defining a probability distribution based on redshift, number of members, projected radius, and velocity dispersion, and then matching this to their observed counterparts. For more information on how these physical properties were found as well as their potential error please see Section~4.2 of K12. For group centers they used a method based on weighting each group member by a Voronoi-Delauney Tessellation, described in Section~4.1 in \citet{presotto_journey_2012}.

Here we limit the analysis to groups in the area covered by the UVCANDELS data in the COSMOS field. In order for a group to be considered ``covered" in the UVCANDELS area, we require that the projected size of the group, taken to be 1.5 times the group radius, falls within the CANDELS/COSMOS field. This selection does include 8 galaxies that have been analyzed in UVCANDELS but were not covered by any UV filters. We have found that excluding these galaxies does not significantly change our results, so we have opted to include them. K12 defined the probability of a group existing, GRP2, based off of whether the group was found using both group finding algorithms they employed. We only consider groups with GRP2 $=1.0$ to ensure we are only looking at groups that confidently exist. We only consider groups with at least 3 spectroscopic members, to ensure the high values of completeness and purity described in K12. The K12 catalog also reports a probability for a galaxy to belong to its host group, computed from the redshift difference and projected distance between the galaxy and the group center. In the analysis below, we only include galaxies that have a probability $\ge 0.8$. Richness was calculated using group members that had a probability of $\ge 0.8$ of being in the group, and was given an upper inclusive limit of 25 members, to avoid contaminating our group sample with cluster galaxies. These cuts yield a  22 group sample across the $0.2 < z < 0.8$ redshift range within the CANDELS/COSMOS field (Figure~\ref{fig:Stamps}). 

\subsection{Color and Mass Complete Sample}

In order to ensure that we have an unbiased mass complete sample, we apply criteria based on redshift, magnitude, and stellar mass. Ideally, a volume limited sample constructed by selecting all galaxies brighter than a rest-frame luminosity would ensure that we are comparing the same objects at all redshifts \citep[e.g.,][]{presotto_journey_2012}. A typical choice in the literature is to use the evolution-corrected rest-frame $B-$band luminosity. The K12 group catalogs inherits the selection function of the parent $z$COSMOS catalog that includes all galaxies with $I_{F814W}$ magnitude brighter than 22.5. The $I_{F814W}$ filter samples the rest-frame $B-$band at $z\sim 0.8$, therefore at this redshift the completeness of a $B-$band selected catalog does not depend on galaxy colors. The resulting catalog, however, would  still be affected by strong mass-dependent biases, as old galaxies of a given stellar mass would preferentially be excluded compared to young galaxies of the same stellar mass, given their higher $M/L$ ratio. Additionally, at redshifts higher (lower) than 0.8, a pure rest-frame $B-$band luminosity selection would preferentially exclude red (blue) galaxies, introducing additional mass-dependent biases in the final sample. In order to limit these effects, we follow \citet{presotto_journey_2012} and define a luminosity and mass limited sample of group and field galaxies as follows. 

First, we use an evolving $B-$band luminosity cut assuming the luminosity evolution from \citet{zucca_zcosmos_2009} where $M^{*}_{Bev} = -20.3 - 5\text{log}(h_{70}) - 1.1z$. This corresponds to a cutoff in $B-$band absolute magnitude of $M_{B,\mathrm{cutoff}} = M^{*}_{Bev} + 0.8$ which we apply for our sample. The derived mass cut follows the same approach as in \citet{iovino_zcosmos_2010}, resulting in a mass cut of $\text{log}(M_{\mathrm{cutoff}}/M_{\odot}) = 10.56$. More details on the methods used to derive the mass and luminosity cuts can be found in \citet{presotto_journey_2012} Section 5.1.
The final volume mass-limited sample of group galaxies includes 19 galaxies in 8 groups in the $0.2 \leq z < 0.45$ redshift range and 34 galaxies in 12 groups in the  $0.45 \leq z < 0.8$ redshift  range, notably removing two entire groups from our sample. 

The comparison sample of field galaxies was selected from the UVCANDELS catalog, applying the same redshift, magnitude and mass cut as for the group galaxy sample. Additionally this field sample was selected to only include galaxies that were detected in the F275W or F435W filters, which cover the UV at these redshifts. These cuts result in 33 and 87 field galaxies in the low and high redshift range, respectively.

\section{Analysis}
\label{sect:analysis}

\begin{figure*}[!ht]
  \begin{center}
    \includegraphics[width=01.0\linewidth]{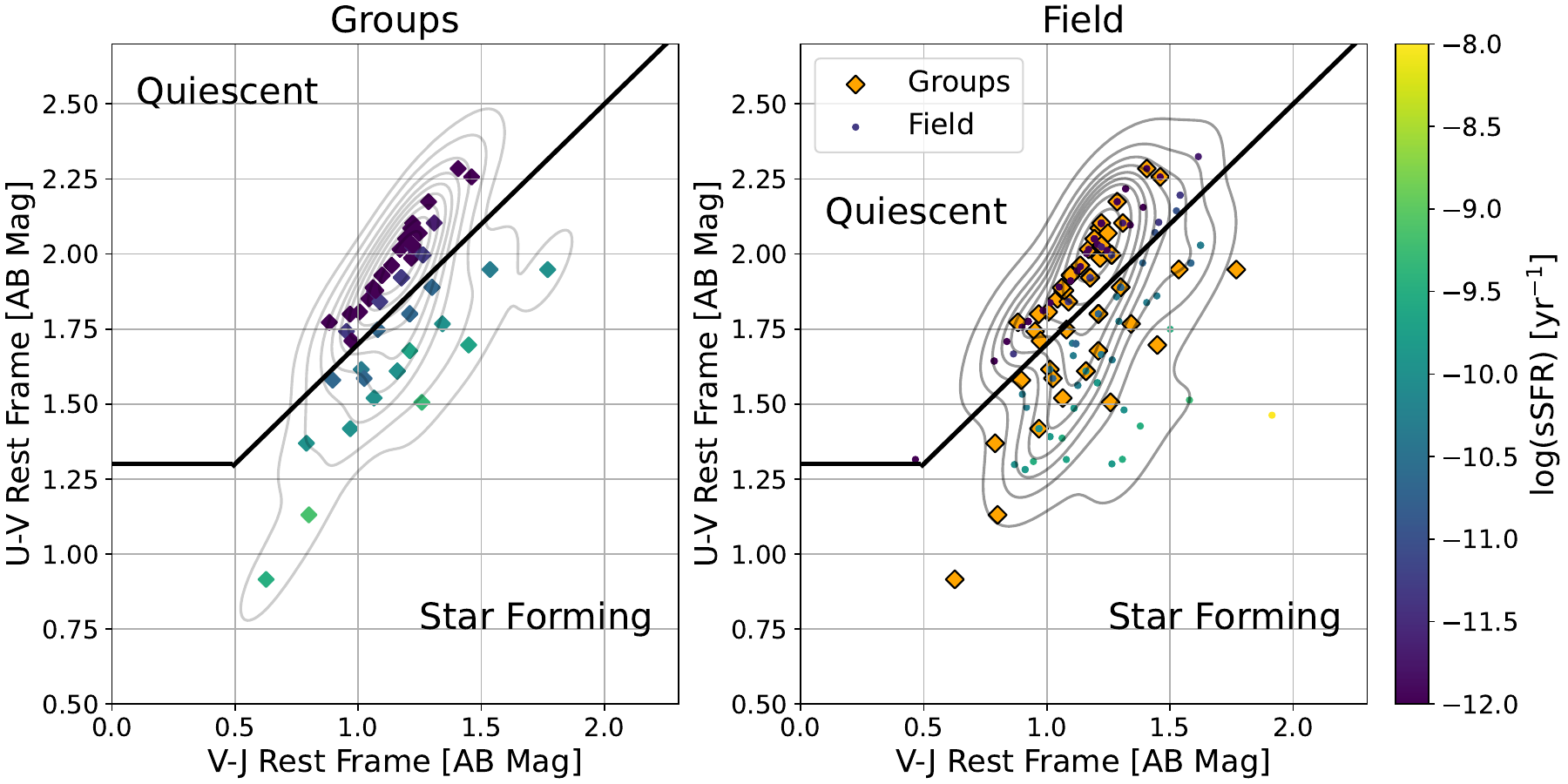}
    \caption{The UVJ diagram used to select the quiescent and star-forming galaxies. The line indicates the cut used to separate quiescent and star-forming galaxies as described in Section~\ref{sec:Qselection}. The colors indicate the log(sSFR) in both the left and right panels. In the left panel we show the group galaxies after the mass and luminosity cuts. The right panel shows the field galaxies from the UVCANDELS COSMOS dataset that are above the $M_{cutoff} = 10.56$ mass cut, absolute $B-$band magnitude $M_{B} < -19.5 -1.1z$ cut, and within the $0.2 < z < 0.8$ range. Overlaid on the right hand side are the group galaxies in orange diamonds. Contour lines are based on the density of galaxies in the UVJ plot.}
    \label{fig:UVJ}
  \end{center}
\end{figure*}

\begin{figure}[!ht]
  \begin{center}
    \includegraphics[width=01.0\linewidth]{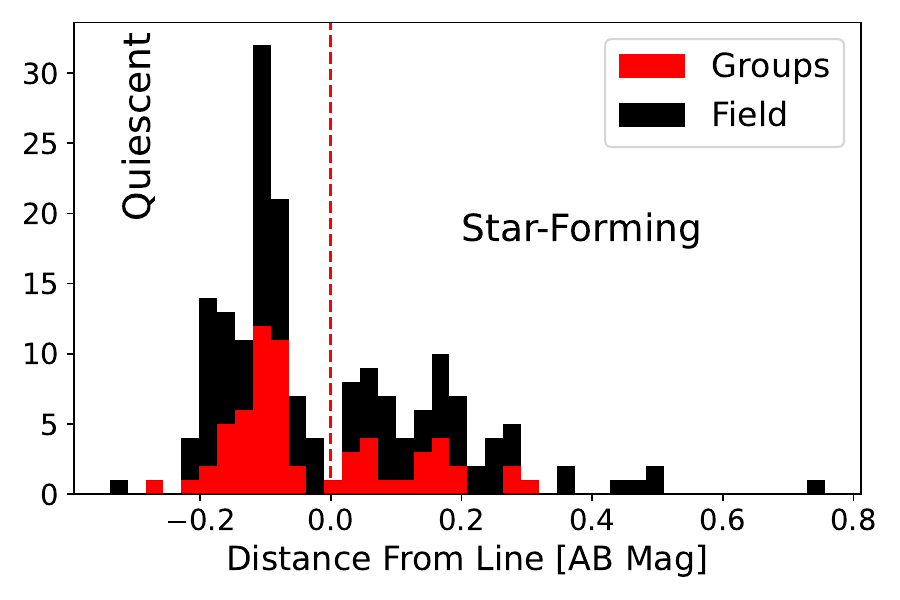}
    \caption{Histogram of distances to the UVJ cut, in AB Mag. The vertical dashed line indicates the UVJ cut that is used to select quiescent galaxies, created through an iterative process described in Section~\ref{sec:Qselection} Negative values are quiescent, positive values are star-forming. Distances were only calculated for galaxies with $U-V>1.3$.}
    \label{fig:UVJDist}
  \end{center}
\end{figure}

\subsection{Selection of quenched galaxies}
\label{sec:Qselection}

The criterion used to separate quenched from star-forming galaxies can influence the quenched fraction, particularly at the high mass end (log(M$_*)\gtrsim~10^{10.5}$M$_{\odot}$). 
\citet{donnari_quenched_2021} demonstrated that there is overall agreement among various definitions  below this threshold. For larger stellar masses, however, different criteria for the identification of passive galaxies results in quenched fractions that can vary between 50-100\%, depending on whether galaxies are centrals or satellites. We discuss why we use the UVJ method and why we believe this minimizes the cross-contamination of the star-forming and quiescent populations in Section~\ref{sec:discussion}.

Here we proceed to use a rest-frame color selection to identify quenched galaxies based on their position in the rest frame $U-V$ vs rest frame $V-J$ diagram  \citep[e.g., hereafter UVJ diagram][]{williams_detection_2009,martis_evolution_2016,whitaker_galaxy_2015}. 
%
%
We use a UVJ color selection because it minimizes contamination from dusty star-forming galaxies. Classifying the data into star-forming and quiescent is necessary not only to analyze the impact of environmental quenching, but also to compare our results to those found in the literature. The UVJ classification has been shown to provide a robust sample of star-forming and quiescent galaxies, minimizing the contamination of either sample \citep{whitaker_newfirm_2011, williams_detection_2009, brammer_number_2011}.

Figure~\ref{fig:UVJ} shows the UVJ diagram for galaxies both in groups and in the field. The  rest--frame colors were computed from the best fit models of the UVCANDELS data  derived with CIGALE (Mehta et al. in prep) using the Johnson U, V, and J bandpasses. Galaxies are color-coded according to their sSFR (SFR/M$_{*}$), as shown in the color-bar on the right hand side of the figure. Quiescent galaxies lie in the cloud on the top left of the distribution, with a lower sSFR that is commonly associated with quiescence \citep{williams_detection_2009, whitaker_star_2012, speagle_highly_2014}. We apply a cut to separate the quiescent galaxies from star-forming galaxies based on the bi-modality of the distribution of colors following the method introduced by \citet{williams_detection_2009}. We first apply a cut to all galaxies with $(U-V)>1.3$\footnote{This cut is applied to remove possible contamination by blue star-forming galaxies.} that visually separates the populations of star-forming and quenched galaxies. We use the field galaxy population, as the large number of galaxies makes the bimodality of the color distribution more prominent.  For each galaxy, we compute the normal distance (in the UVJ plane) to the diagonal dividing line and plot the histogram of the distances in Figure~\ref{fig:UVJDist}. The final parameters (intercept and slope) of the separating line are determined with an iterative method. First, we adjust the slope of the cut to maximize the bi-modality in Figure \ref{fig:UVJDist}, and then we modify the intercept of the line to lie at the minimum between the two peaks. This method converged within two iterations, resulting in the definition of a quiescent galaxy:

\begin{equation}
\begin{split}
   (U-V) \geq 0.8(V-J) + 0.9;\\
   (U-V) > 1.33.
\end{split}
\end{equation}

This cut is in general agreement with the literature values reported by \citet{williams_detection_2009} and \citet{whitaker_galaxy_2015}. Our cut is slightly higher, \citet{williams_detection_2009} reports a slope of 0.88 and intercept of 0.69 while \citet{whitaker_galaxy_2015} uses a slope of 0.8 and intercept of 0.7. These discrepancies are likely due to differences in the bandpass definition used to compute the U-V, and V-J rest-frame colors in our analysis compared to others in the literature. Note that we do not consider a redshift-dependent selection of quiescent galaxies. \citet{williams_detection_2009} find that the $U-V$ color for passive galaxies evolves by less than 0.15 magnitudes out to $z\sim 2$, and thus this effect is negligible in the redshift range considered here. The log(sSFR), indicated by the color, of the galaxies in Figure~\ref{fig:UVJ} confirms that in the selected region the majority of galaxies have low sSFR (log(sSFR)$<-11$) associated with quenched  systems.

\subsection{The quenched fraction}
\label{sec:quenchedFraction}

In the following we compute the probability of a galaxy being quenched, $f_Q$ and $\widehat{f_Q}$. When calculating how the probability of being quenched depends upon a given predictor (redshift, group-center distance, or stellar mass) we use a logistic regression model as it helps us avoid binning of the data. In using this model we assume that $f_Q$, the probability of being quenched, follows a logistic function, with the predictor as the input variable x.
\begin{equation}
\displaystyle f_Q = \frac{1}{1+e^{-(bx+a)}}
\end{equation}

To compute the a and b parameters of this model we utilize Bayes' Theorem, implementing an MCMC algorithm with the \textsc{pymc3} python package \citep{salvatier_pymc3_2016}, using a normal distribution with a mean of 0 and standard deviation of 10 for the priors on a and b. With 5000 steps and the NUTS sampler we are able to find the maximum a posteriori for a and b, listed for our two models below. Due to the multivariate nature of logistic regression we were able to fit two models, one that takes redshift and stellar mass to derive the mass dependence and redshift evolution of $f_Q$ (Figures~\ref{fig:fQvsZ}~and~\ref{fig:fQvsMass}) and a second model that takes redshift, stellar mass, group halo mass and normalized distance to the group center, normalized by the radius of the group. The Maximum a Posteriori for this first model are $a = -9.03^{+8.7}_{-8.7}$, $b_{z} = -2.2^{+1.8}_{-1.8}$, and $b_{log(M)} = 1.021^{+0.8}_{-0.8}$ for the group model and  $a = -8.35^{+7.0}_{-6.9}$, $b_{z} = -0.35^{+1.1}_{-1.1}$, and $b_{log(M)} = 0.81^{+0.64}_{-0.64}$ for the field model, where $b_z$ is the slope for the redshift, and $b_{logM}$ is the slope for the stellar mass. The Maximum a Posteriori for the second model is $a = -0.38^{+9.9}_{-9.4}$, $b_{z} = 7.49^{+9.3}_{-9.4}$, $b_{logM} = 3.32^{+1.9}_{-1.8}$, $b_r = ^{+1.5}_{-1.5}$, and $b_{GM} = -3.85^{+2.7}_{-2.76}$, where $b_z$ is the slope for the redshift, $b_{r}$ is the slope for the normalized radius, $b_{logM}$ is the slope for the galaxy stellar mass, and $b_{GM}$ is the slope for the group halo mass. This second model is used to analyze the radial distribution of $f_Q$ in Figure~\ref{fig:Quenched}. For each model we marginalize over the variables not being plotted.

To ensure our models best reflect the trends in the data we have chosen uninformative priors for a and b. To verify that our results were not sensitive to small changes in these priors we performed a sensitivity analysis considering a range of normal priors (mean from -5 to 5 and standard deviation from 1 to 10). This analysis shows that, unless a very small standard deviation for the priors is used, the resulting posterior does not change substantially. Therefore our results are not significantly dependent to  changes to the chosen priors.

In the side panels of Figure \ref{fig:Quenched} we define $\widehat{f_Q}$, to provide a general estimate of $f_Q$ averaged over the redshift bin, group centric radius, and stellar mass. We define $\widehat{f_Q}$ as the number of quenched galaxies ($n_q$) divided by the total number of galaxies ($n_T$) in that bin. To calculate the error on $\widehat{f_Q}$ we use Bayes' theorem. Assuming that the likelihood of observing $n_Q$ quenched galaxies out of a sample of $n_T$ is Binomial with probability $p = \widehat{f_Q}$ of a galaxy being quenched with $k = n_q$ successes and $n = n_T$ trials, then the natural choice for a prior is a Beta distribution with parameters $\alpha=\beta=1$. In each considered bin, we report the maximum a posteriori value of $\widehat{f_Q}$ and the 68\% confidence interval.

\begin{equation}
    \text{Beta}(\alpha,\beta) \propto x^{\alpha - 1}(1-x)^{\beta - 1}
\end{equation}

\begin{equation}
    \text{Binom}(k,n,p) \propto p^k(1-p)^{n+k}
\end{equation}

\section{Results }
\label{sect:results}

\begin{figure}[!ht]
  \begin{center}
    \includegraphics[width=01.0\linewidth]{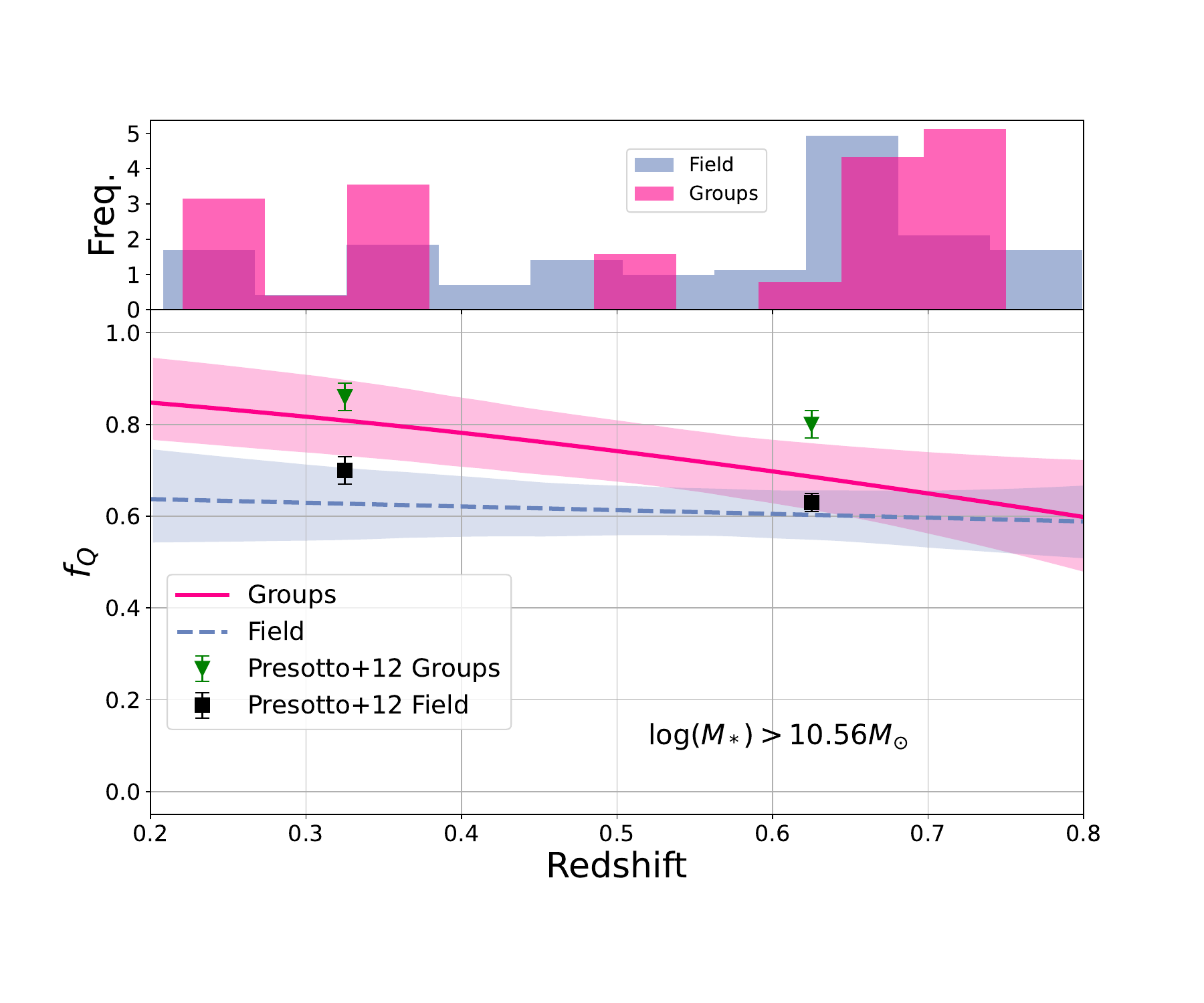}
    \caption{The probability of being quenched ($f_Q$) derived from a logistic regression model versus the redshift of the galaxy. Only galaxies with $M_B < -19.5 - 1.1z$ and log$(M_*/M_{\odot}) > 10.56$ are included. The red solid line corresponds to the logistic regression of the group galaxies, while the dashed black line corresponds to the logistic regression of the field galaxies. The shaded regions correspond to the 68\% percentile for each regression. Green upside-down triangles denote the field galaxies from \citet{presotto_journey_2012}, black squares denote the group galaxies from \citet{presotto_journey_2012}. The gradual deviation of group galaxies from field galaxies suggests the group environment slowly changes the probability of galaxies being quenched, implying a slow quenching process is dominant in the group environment.}
    \label{fig:fQvsZ}
  \end{center}
\end{figure}

For a rough understanding of the timescale of evolution for $f_Q$ of group and field galaxies we look at the redshift evolution of $f_Q$ in Figure~\ref{fig:fQvsZ}. In this figure we show the redshift evolution of the quenched fraction of galaxies with a stellar mass greater than $10^{10.56}$M$_{\odot}$ and with absolute $B-$band magnitude $M_{B} < -19.5 -1.1z$. The red and black lines show the Bayesian estimate of $f_Q$ together with with the 68\% confidence interval. The trend of the field is plotted in a black dashed line, while the trend of the groups is plotted in a solid red line. 
The top panel displays the distribution of field and group galaxies in redshift space. We find that $f_Q$ for group galaxies increases as redshift decreases, from \highZgroups at $z=0.8$ to \lowZgroups at $z=0.2$. For field galaxies our results suggest that the quenched fraction remains relatively constant as a function of redshift, with  $f_Q$ varying between \highZfield to \lowZfield. This result is in agreement with values reported in the literature \citep[e.g.,][]{knobel_colors_2013,peng_mass_2010, presotto_journey_2012, ji_evidence_2018, donnari_quenched_2020}. For example, \citet{presotto_journey_2012} find that the fraction of quenched galaxies among massive objects in groups increases from 0.8$^{+0.03}_{-0.03}$ to 0.86$^{+0.03}_{-0.03}$ between redshifts 0.65 and 0.3, respectively. 

Figure~\ref{fig:fQvsZ} shows that the difference between field and group galaxies becomes stronger as redshift decreases. The overall trend of the group galaxies compared to the stagnant field suggests the group environment significantly influences a galaxy's probability of being quenched. This influence either occurs the longer a galaxy is in a group, or in groups that have formed at redshifts around 0.325. 

The distributions of group and field galaxies illustrated in Figure~\ref{fig:fQvsZ} are only separated at lower redshifts and only by a small amount. To verify if these field and group distributions are unique we perform three Anderson-Darling tests. The first test is over the entire redshift range, the second and third are over the higher ($0.45 < z < 0.8$) and lower ($0.2 < z < 0.45$) redshift ranges, to see how the difference between the group and field distributions change over time. The p-value the Anderson-Darling test over the entire redshift range is 0.08. Typically two distributions are only considered significantly different if the p-value is less than 0.05, indicating over the entire redshift range the two distributions could be drawn from the same sample. The p-value of the upper redshift range is $>0.25$, indicating the two distributions are likely drawn from the same distribution. The p-value in the lower redshift range is $0.02$, meaning in the lower redshift range the two distributions are confidently different. What these Anderson-Darling tests tell us is that while the field and group distributions do not confidently differ over the entire redshift range they begin to differ significantly at lower redshifts.

\begin{figure}[ht]
  \begin{center}
    \includegraphics[width=\linewidth]{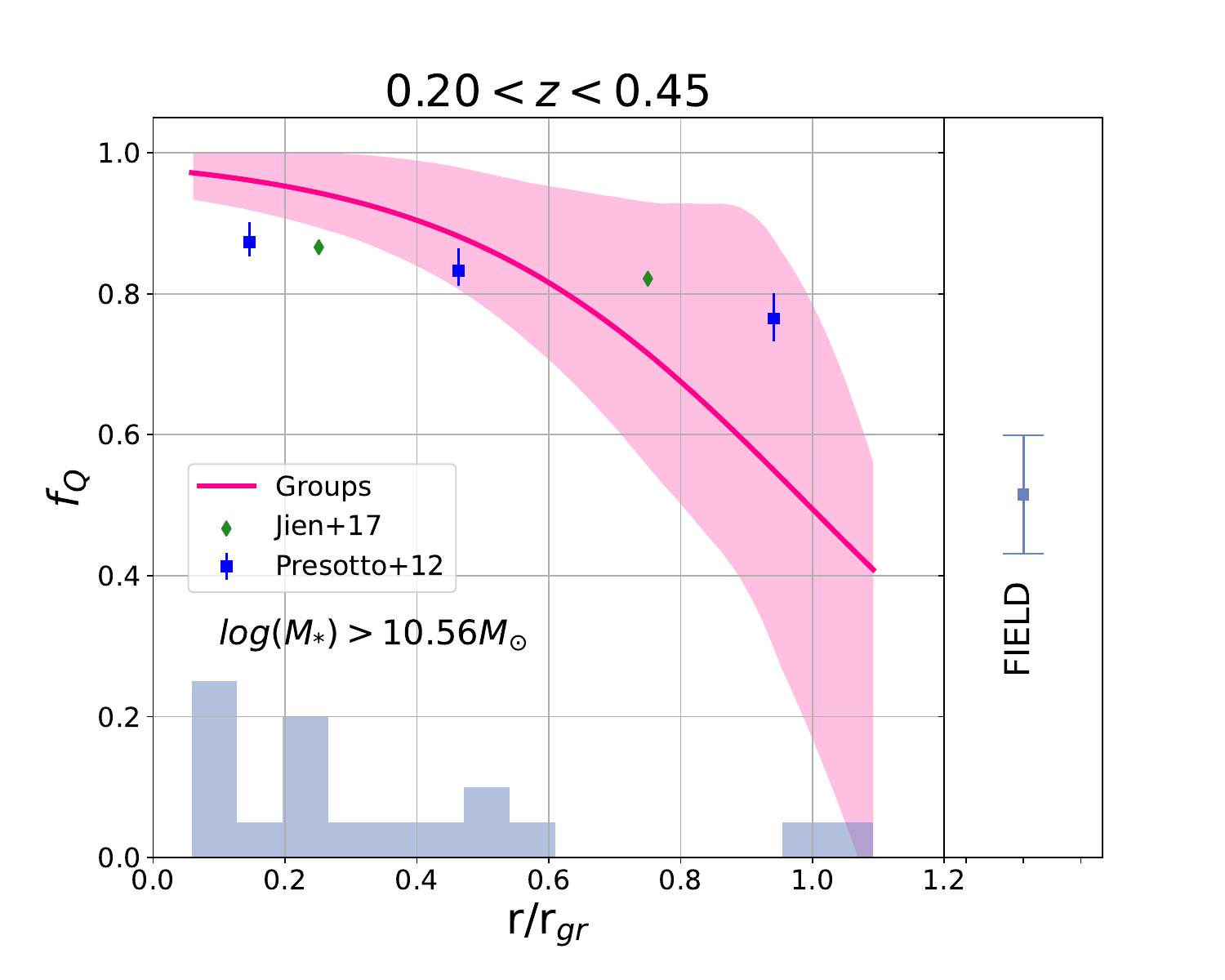}
    \caption{The probability of being quenched($f_Q$) derived from a logistic regression model versus the distance to the group center ($r/r_{gr}$) normalized by the radius of the group ($r_{gr}$). The histogram indicates the normalized distribution of group galaxies that went into constructing the model.
    The galaxies are limited to a redshift range of $0.2 < z < 0.45$, stellar mass of log$(M_*/M_{\odot}) \geq 10.56$, and B-band absolute magnitude of $M_{B} < -19.5 -1.1z$. The shaded regions correspond to the 68\% interval of the regression model. The right hand sidebar  indicates $\widehat{f_Q}$ for the field galaxies within this redshift range. Blue boxes denote the galaxies from \citet{presotto_journey_2012}, while the green diamonds denote the galaxies from \citet{jian_pan-starrs1_2017}. The radial trend suggests that the probability of being quenched slowly increases with the time galaxies spend in groups, indicative of a slow quenching process.}
    \label{fig:Quenched}
  \end{center}
\end{figure}

To investigate the timescale of environmental quenching within each redshift bin we explore the radial dependence of $f_Q$ in Figure~\ref{fig:Quenched}. We present the Bayesian estimate of $f_Q$, with the 68\% interval, as a function of the distance to the group center ($r/r_{gr}$) normalized by the radius of the group ($r_{gr}$), derived in K12. The histogram indicats the radial distance distribution of galaxies used to derive the model. Finally, in the side panel, we show the value of $\widehat{f_Q}$ for field galaxies. 

Although the uncertainties are large we do see a general radial dependence of $f_Q$. In the outskirt of groups, $f_Q = $\lowZhighR, consistent with the average value measured in field galaxies ($f_Q = $\lowZfieldAve) in the same redshift bin. Moving toward the central parts of groups, $f_Q$ increases, reaching values of $f_Q = $\lowZlowR. When interpreting this trend we must be careful, as beyond $r/r_{gr} = 0.6$ we only have two points constraining our fit, which must be considered when we present our analysis in Section~\ref{sect:analysis}. However, the general trend indicates that the probability of being quenched is loosely dependent on the distance a agalxy is to its group center.

Figure~\ref{fig:Quenched} only includes the lower redshift bin as the upper bin of $0.45 < z < 0.8$ is subject to differing uncertainties that make its interpretation difficult. We analyze the radial trend of $f_Q$ with the assumption that $r/r_{gr}$ is statistically correlated with the time since infall \citep{taranu_quenching_2014, gao_subhalo_2004}. With the increased likelihood of non-\-virialized halos and contamination at higher redshifts this assumption breaks down. So we do not analyze the radial trend of $f_Q$ in the $0.45 < z < 0.8$ redshift bin.

\begin{figure*}[ht!]
  \begin{center}
    \includegraphics[width=\linewidth]{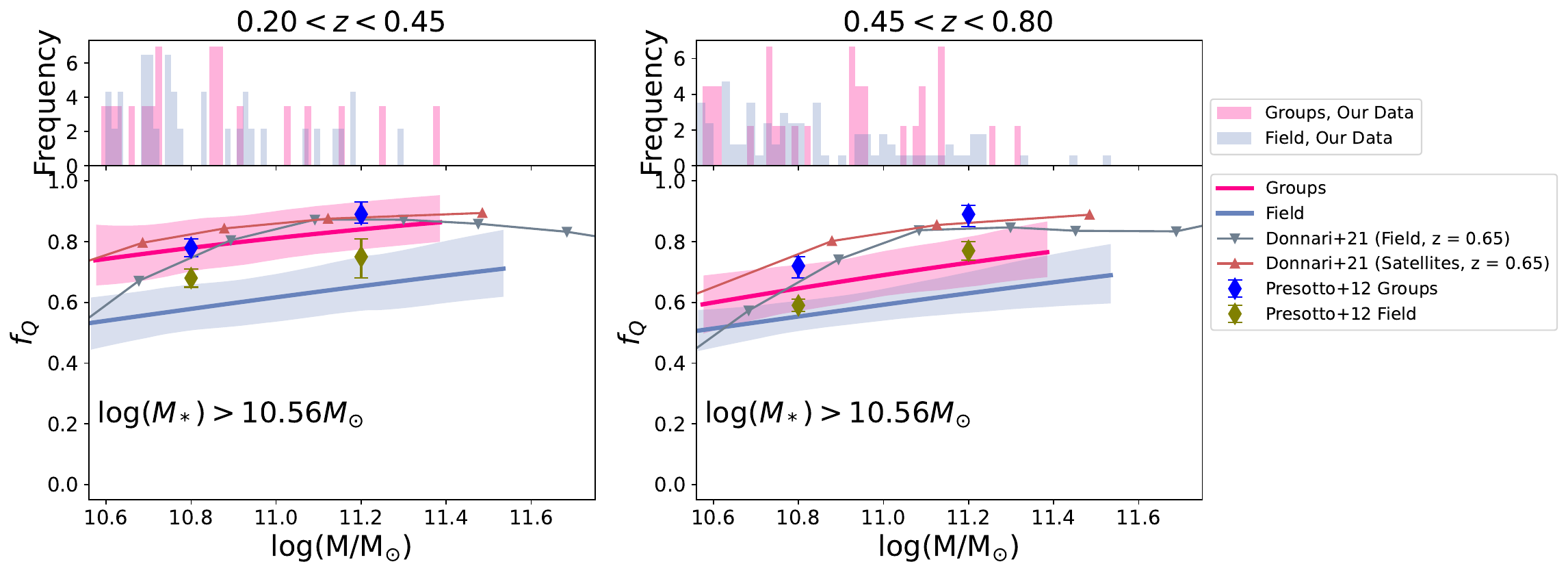}
    \caption{The bottom panels show the probability of being quenched ($f_Q$) derived from a logistic regression model versus the stellar mass of the galaxies. The solid red line corresponds to the group galaxies, while the black dashed line corresponds to the field galaxies. The left hand side is group galaxies within the $0.2 < z < 0.45$ range, the right hand plot is group galaxies within the $0.45 < z < 0.8$ range, both are limited to log$(M_*/M_{\odot}) \geq 10.56$ and $M_{B} < -19.5 -1.1z$. The shaded regions are the 68\% and 95\% percentiles from the regression models. The top two plots are weighted histograms of the group and field mass distributions, weighted so the area under their histograms integrates to 1. Upside-\-down grey triangles denote the simulated field values from \citet{donnari_quenched_2020}, while the red triangles denote the simulated satellite galaxies from \citet{donnari_quenched_2020}. The blue diamonds indicate the group values from \citet{presotto_journey_2012}, while the green diamonds denote the field values from \citet{presotto_journey_2012}. Both panels show that galaxies of higher stellar masses are more likely to undergo quenching. However, the effects of the environment become more apparent in the left-hand panel, where the $f_Q$ of group galaxies does not exhibit a clear trend with mass, unlike the field at the same redshift. We also see in the top panels that the mass distribution of group and field galaxies are similar.}
    \label{fig:fQvsMass}
    \end{center}
\end{figure*}

To explore the possibility of the trend in Figure \ref{fig:Quenched} being driven by mass quenching, in Figure~\ref{fig:fQvsMass} we investigate how the probability of being quenched changes with galaxy stellar mass. As in the previous figures, we show the logistic regression of $f_Q$ and the corresponding 68\% confidence intervals. The low and high redshift bins are shown on the left and right panels, respectively. The top panel in each shows the mass distributions of galaxies in groups and in the field.

The probability of being quenched depends in similar ways on the stellar mass for group and field galaxies for the $0.45 < z < 0.8$ redshift range. Specifically, $f_Q$ increases with with increasing stellar mass amoung the group and field galaxies equally. This trend changes in the $0.2 < z < 0.45$ redshift range, where $f_Q$ is similarly dependent on mass in the field but seems independent of mass in groups. We find that in this lower redshift range the mass dependence of the groups and field are confidently different, with an Anderson-Darling p-value of 0.017.
These relationships are in relative agreement with the literature \citep{wetzel_galaxy_2013, presotto_journey_2012, ji_evidence_2018} with our values of $f_Q$ being within the standard deviation of their results. 

Due to the mass dependence of $f_Q$, its {\it radial} and {\it redshift} dependencies could be introduced if the mass distribution changes as a function of distance from the center of the group. We have tested this possibility and found that, at all redshifts, there is no correlation between a galaxy's stellar mass and the distance to its group's center nor its redshift. 

We have refrained from analyzing the relationship between $f_Q$ and the group halo mass due to the size of our group sample. We have a total of 8 group in our lower redshift sample and 12 in our upper redshift sample. We do not believe that we can conduct a robust analysis on how $f_Q$ changes with just the group halo mass. In an effort to marginalize over the effects that the group halo mass has on our results we have included the group halo mass in the presented in Figure~\ref{fig:Quenched}, where we have marginalized over the group halo mass.

\section{Discussion}
\label{sec:discussion}
We begin the discussion of the results presented in the previous sections by addressing possible sources of systematic uncertainties, including the comparison with different techniques used in similar studies in the literature and then discuss the quenching timescales.

\subsection{Systematic uncertainties}
\label{sec:systematics}

The UVCANDELS dataset covers only approximately 4\% of the COSMOS field used in K12, limiting the number of groups for which the new UV data is available. Small number statistics could then cause some of the trends we observe as well as add uncertainty in the measured values. However, the relative agreement that we observe between our trends and those resulting from larger samples of \citet{presotto_journey_2012} and \citet[][]{ji_evidence_2018} suggests that the effect of small-number statistics is limited  to increasing the uncertainties in the results and does not drive the trends we observed.

Another source of systematic uncertainty is caused by our selection method of quenched galaxies. It has been shown that the strategy used to identify the sample of quiescent galaxies can introduce significant differences in the measured values of $f_Q$, particularly for massive galaxies, with log(M$_*$/M$_{\odot}) > 10.5$ \citep{donnari_quenched_2021}. We therefore explored how our choice of using the UVJ diagram, as opposed to other definitions in the literature, influences our results. To test the severity of this effect we compare three definitions of quiescence.  Specifically, we compare the UVJ methods from \citet{williams_detection_2009} and this work, the rest-frame $U-B$ vs stellar mass method used in \citet{presotto_journey_2012}, as well as two differing sSFR (SFR/Mass) thresholds from \citet{jian_pan-starrs1_2017} and \citet{donnari_quenched_2020}. We find that using either the two UVJ methods or the \citet{donnari_quenched_2020} sSFR method does not change the redshift evolution of the group and field galaxies significantly nor the radial trend of $f_Q$. However, we find that using the methods from \citet{presotto_journey_2012} and \citet{jian_pan-starrs1_2017} give us significantly differing redshift evolutions and radial trends of $f_Q$. We believe that the sSFR threshold definition in \citet{jian_pan-starrs1_2017} is too high and allows for star forming galaxies to contaminate the quiescent population at high masses. This is due to the turn off of the SFR-M main sequence \citep{speagle_highly_2014}, where galaxies with a lower sSFR are still characterized as star-forming. The U-B vs stellar mass definition used in \citet{presotto_journey_2012} also suffers from a contaminated quiescent population. We find that this method fails to completely disentangle dusty star-forming galaxies from quiescent galaxies, unlike the UVJ method used here and in \citet{williams_detection_2009}. While contamination is always likely with any definition of quiescence we believe that the robustness of our measurements when using our UVJ definition, that of \citet{williams_detection_2009} and the sSFR definition of \citet{donnari_quenched_2020} suggests we have minimized the contamination in our sample.

\citet{presotto_journey_2012}  studied the entire 20k zCOSMOS group catalog, with similar mass, luminosity, and redshift cuts as we placed on our data. While they did not focus on the dominant quenching mechanism, they found similar radial trends and redshift evolution of $f_Q$.  Their values of $f_Q$ are consistently higher than what we find. This difference can be explained by their adopted selection of quenched galaxies. Specifically,  they define quiescence using the $U-B$ versus stellar mass diagram. As Figure~\ref{fig:UVJ} shows, a cut based on only one color will result in a sample that includes star-forming dusty galaxies, increasing the measured $f_Q$. Despite the value of $f_Q$ being larger, however, the observed trends with redshift and group properties are similar to those we find here (see, e.g., Figure~\ref{fig:fQvsZ}).

Another recent work addressing the evolution of the quenched fraction as a function of environment is  \citet{jian_pan-starrs1_2017}, who use the Pan-STARRS1 medium-deep survey \citep{kaiser_pan-starrs_2010} and focus on the same redshift range as our analysis. Their quenched fraction are higher than our values, in this case because the sSFR threshold they use in their analysis (10$^{-10.1}$ yr$^{-1}$) includes galaxies that we consider star-forming (e.g., Figure~\ref{fig:UVJ}). Furthermore their analysis is dependent on the radial trend of high mass galaxies, where the contamination of star-forming galaxies becomes even more critical due to the turn-off of the SFR-M main sequence. 
However, comparisons with their radial trends is not ideal as they do not provide normalized distances to the group center, nor the group radii. To allow for any comparison we normalize their radial distances by 1.5~Mpc, which they quote as the average radius of their groups. It is because of these reasons that while our results are withing agreement of their own we come to differing conclusion, as our analysis is not contaminated by star-forming galaxies to the extent that theirs is.

An additional uncertainty can be introduced by the assumption that the group dark matter halos are fully virialized. This assumption is used in the creation of the group catalog by K12, but it also enters in the analysis below when we link the group-center distance to the time since infall. This assumption may break for the highest redshift bin considered  ($z > 0.45$). Groups' halos not being fully virialized could introduce an ambiguity in the definition of their radii, possibly resulting in incomplete (galaxies which are entering the group would be missed by our selection) and contaminated group catalogs  \citep{alpaslan_galaxy_2012}. However, we expect this problem  to be minimized in our analysis as we limit our sample to groups with the highest estimated purity and completeness.

\subsection{Quenching processes}

Environmental quenching processes can be broadly separated into rapid and slow mechanisms depending on whether they remove the entirety of the cold material in a galaxy, or only act on the outermost, low-density gas.  
In Section~\ref{sect:results} we presented how the probability of being quenched ($f_Q$) for group galaxies depends on redshift, distance to the group center, and stellar mass. Here we analyze these trends to try and determine whether environmental quenching in groups is dominated by a slow or rapid mechanism.

Figure~\ref{fig:fQvsZ} shows that the redshift evolution of $f_Q$ evolves differently for galaxies in groups than in the field. The gradual deviation between group and field galaxies suggests that the dominant quenching process in these groups is slow. If the dominant process is rapid then we expect $f_Q$ in groups to significantly differ from the field at all redshifts. Seeing the redshift evolution gradually diverge suggests that a majority of group galaxies are quenched only after several billion years after their accretion into a group. Direct interpretation of the redshift dependency of $f_Q$, however, can be blurred by the fact that groups continue to accrete galaxies from the field environment, thus mixing together the quenching and accretion processes. This effect would be more important for slow quenching mechanisms (acting on timescales similar to those of accreting new galaxies) and  minimized for rapid mechanisms (of the order of few Myrs, e.g., for ram pressure stripping).

A clearer picture of the influence of the group environment on quenching can be obtained by also looking at the distribution of $f_Q$ as a function of how long group galaxies have been in groups. This can be done by examining the radial dependence of $f_Q$, because the projected distance to the group center ($r/r_{gr}$) is found to statistically correlate with the time since a galaxy's infall in the group halo. Numerical simulations show that the infall lookback time slowly declines with $r/r_{gr}$, indicating that, statistically, galaxies projected closer to the center of groups have been in the group for a longer time \citep{taranu_quenching_2014, gao_subhalo_2004}. 

With this in mind, Figure~\ref{fig:fQvsZ} shows that, in the lower redshift bin\footnote{We refrain from commenting on the high redshift bin because, as we discussed in \ref{sec:systematics}, we believe that these groups are more likely to not yet be fully virialized.}, the longer a galaxy has been inside a group the more likely it is to be quenched, confirming  that the group environment influences the probability of being quenched. Interpreting this result in terms of quenching processes becomes difficult due to the low number of points constraining the outer values of our fit. If we observe the trend starting a $r/r_{gr} \leq 0.6$, where we are more constrained, we see that $f_Q$ is still within agreement of the field galaxies. This suggests that even when galaxies have been in a group for $> 1$ Gyr their likelihood of being quenching is consistent with that of a field galaxy. Combined with the slow redshift evolution this radial trend suggests that the dominant environmental quenching process acts on timescales of several billion years.

We emphasize that while our best estimate for the timescale of environmental quenching is long, implying a slow quenching process, more data are needed to confirm this result. Indeed, while the best fit of our models were used to conduct the analysis, the standard deviation is large, and thus we cannot completely rule out rapid quenching as a dominant mechanism in groups.

To quantify the quenching timescale ($t_Q$) of our lower redshift bin, we follow the work of \citet{foltz_evolution_2018}, who proposed a method for determining $t_Q$ based on  connecting the dark-matter mass accretion rates to the observed numbers of quenched and star-forming galaxies. To briefly summarize, we first define the mass accretion rate of the group dark matter halo;

\begin{equation}
\begin{split}
    \frac{dM}{dt} = 46.1 \text{M}_{\odot}\text{yr}^{-1} \left(\frac{M}{10^{12}M_{\odot}}\right)^{1.1} \\
    *\left(1+1.11z\right)\sqrt{\Omega_m(1+z)^3+\Omega_{\lambda}}
\end{split}
\end{equation}

\begin{equation}
\begin{split}
    M(z_c) = M_0
\end{split}
\end{equation}

\noindent Where M is the mass of the group dark matter halo, M$_0$ is the observed mass of the halo (calculated in K12), $\Omega_m$ is the matter overdensity of the universe, and $\Omega_{\lambda}$ is the dark energy overdensity of the universe. To solve this differential equation we use the observed redshift, $z_c$ and mass, $M_0$ of a given group as the initial value. To solve for the quenching time we assume that all star-forming galaxies (B) have been in their group for less than the quenching time, while all quenched (R) galaxies have been in the group for at least the quenching time. From this we can relate the ratio of B and R to the ratio of the group dark matter halo mass at the time it is observed (M($z_c$)) and one quenching time before it was observed (M($z_c + \Delta z_Q$));

\begin{equation}
\begin{split}
    \frac{B}{R} = \frac{M(z_c) - M(z_c + \Delta z_Q)}{M(z_c + \Delta z_Q)}
\end{split}
\end{equation}

\noindent For a more detailed explanation of the reasoning we direct the reader's attention to Appendix~B of \citet{foltz_evolution_2018}.

To calculate the error on our derived $t_Q$ we use a Monte Carlo simulation where we vary the ratio of B and R as well as $z_c$ and $M_0$. In each realization of the simulation we select $z_c$ and $M_0$ from the distribution of group dark matter halo masses in our sample. We also derive B/R from the distribution of $\widehat{f_Q}$ for the group galaxies in this redshift range. We use the spread of the resulting $t_Q$ distribution as an estimate of the systematic uncertainty associated with $t_Q$. We find a quenching timescale  $t_Q = 4.91^{+0.91}_{-1.47} $Gyrs. This method, however, assumes a slow timescale and depends on the exact value of $f_Q$, which in turn, as we showed in the previous section, depends on the definition used to select quenched galaxies.

\section{Conclusions}
\label{sect:con}

We use a sample of 20 groups from the zCOSMOS 20k group catalog in the redshift range of $0.2 < z < 0.8$ to study the dominant method of environmental quenching in galaxy groups. We added new UV data from UVCANDELS to existing multiband photometry to derive galaxy physical properties. 

We identify quenched galaxies using the UVJ diagram and explore the probability of being quenched ($f_Q$) in groups and the field as a function of redshift, stellar mass, and time within a group. 
Limiting the analysis to galaxies with log$(M_*/M_{\odot}) > 10.56$, we find that $f_Q$ changes slowly with redshift, from \highZgroups at $z=0.8$ to \lowZgroups at $z = 0.2$, compared to field galaxies which remains constant around \lowZfield from $z = 0.8$ to $z = 0.2$. We find that  $f_Q$ decreases as a function of distance to the group center in the lower redshift bin, suggesting that the longer a galaxy has been in a group the higher the probability of it being quenched. Combined the radial trend of $f_Q$ and the redshift evolution of $f_Q$ suggests that the dominent environmental quenching mechanism in our sample is a slow process, such as strangulation \citep{larson_evolution_1980} or delayed-then rapid \citep{wetzel_galaxy_2013}. Assuming the quenching process is slow we used the fraction of quenched galaxies to compute quenching timescales of $t_Q = 4.91^{+0.91}_{-1.47} $Gyrs. This number, however, is highly uncertain because of the assumed halo mass accretion rate and its dependency on the definition of quenched.  

In future work, we plan to extend the analysis of the impact of group environment on galactic star-formation to the remaining three fields that are covered by the UVCANDELS survey. The UV and B-band data available in this survey allows for improved photometric redshift quality, and thus have a more robust field sample. Future exploration of a wider array of environments such as with the upcoming Euclid Space Telescope, Nancy Grace Roman Space Telescope, and recently launched JWST will help further constrain the role of the environment within groups. Fields such as COSMOS, with the wide array of band coverage, create a unique data-set to study these environmental effects, and future UV projects in this field would offer even further constraints on the vital role of environment on extragalactic evolution.

\section*{Acknowledgements}
Based on observations with the NASA/ESA/CSA Hubble Space Telescope obtained at the Space Telescope Science Institute, which is operated by the Association of Universities for Research in Astronomy, Incorporated, under NASA contract NAS5- 26555. Support for Program number 15647 was provided through a grant from the STScI under NASA contract NAS5-26555.

Y. Sophia Dai acknowledges the support from National Key R\&D Program of China via grant No.2017YFA0402704, the NSFC grants 11933003, and the China Manned Space Project with No. CMS-CSST-2021-A05.

L. Y. Aaron Yung is supported by an appointment to the NASA Postdoctoral Program (NPP) at NASA Goddard Space Flight Center, administered by Oak Ridge Associated Universities under contract with NASA.

Rogier A. Windhorst acknowledges support from NASA JWST Interdisciplinary Scientist
grants NAG5-12460, NNX14AN10G and 80NSSC18K0200 from GSFC.

Special thanks to Prof. Galin Jones for helping us revise our statistical analysis.

\bibliographystyle{aas.bxt}
\bibliography{main.bib}

\end{document}